\documentclass[prl,twocolumn,showpacs,preprintnumbers,amsmath,amssymb,floatfix]{revtex4}
\usepackage{graphicx}

\def\tHp{\tau_{\rm Hp}}

\def\Shp{S_{\rm Hp}}

\def\rsa{r_{\rm s1}}
\def\rsb{r_{\rm s2}}
\def\rsc{r_{\rm s3}}

\begin{document}

\preprint{physics/0412046}

\title{Nonlinear Evolution of $q=1$ Triple Tearing Modes in a Tokamak
Plasma}

\author{
Andreas~Bierwage$^{1}$\footnote{e-mail:bierwage@center.iae.kyoto-u.ac.jp},
Satoshi~Hamaguchi$^{2}$\footnote{e-mail:hamaguch@ppl.eng.osaka-u.ac.jp},
Masahiro~Wakatani$^{1}$\footnote{deceased},
Sadruddin~Benkadda$^{3}$,
Xavier~Leoncini$^{3}$}
\affiliation{
$^{1}$ Graduate School of Energy Science, Kyoto University, Gokasho,
Uji, Kyoto 611-0011, Japan \\
$^{2}$ STAMIC, Graduate School of Engineering, Osaka University, 2-1
Yamadaoka, Suita, Osaka 565-0871, Japan \\
$^{3}$ PIIM-UMR 6633 CNRS-Universit\'{e} de Provence, Centre
Universitaire de St J\'{e}r\^{o}me, case 321, 13397 Marseilles Cedex 20,
France}

\date{\today}

\begin{abstract}
In magnetic configurations with two or three $q=1$ (with $q$ being the
safety factor) resonant surfaces in a tokamak plasma, resistive
magnetohydrodynamic modes with poloidal mode numbers $m$ much larger
than 1 are found to be linearly unstable. It is found that these
high-$m$ double or triple tearing modes significantly enhance through
nonlinear interactions, the growth of the $m=1$ mode. This may account
for the sudden onset of the internal resistive kink, i.e., the fast
sawtooth trigger. Based on the subsequent reconnection dynamics that
can proceed without formation of the $m=1$ islands, it is proposed
that high-$m$ triple tearing modes are a possible mechanism for
precursor-free partial collapses during sawtooth oscillations.
\end{abstract}


\maketitle

\thispagestyle{empty}

The profile of the safety factor $q(r)$ (which measures the magnetic
field line pitch) contains information about the instability
characteristics of magnetically confined plasmas in toroidal or
helical systems with respect to current-driven magnetohydrodynamic
(MHD) instabilities. In particular, in the presence of magnetic
surfaces where $q=1$, a mixing of the plasma inside these surfaces was
observed. This instability is thought to be closely related to
internal disruptions, generally known as sawtooth oscillations, which
strongly affect the quality of energy and particle confinement. In
view of the desired application to thermonuclear fusion reactors such
as ITER, a detailed understanding of these internal large-scale
instabilities is necessary.

A heuristic model proposed by Kadomtsev \cite{Kadomtsev75}
successfully explains overall phenomena associated with a full
sawtooth crash. In this model, a perturbation with helicity $h=m/n=1$
($m$ being the poloidal and $n$ the toroidal Fourier mode number),
which is in resonance with the closed field lines on the $q=1$
surface, quenches the hot core region inside the $q=1$ surface through
magnetic reconnection. This kind of relaxation, generally known as the
$m=1$ internal resistive kink instability (in short, the $m=1$ mode) has
been observed to exhibit an abrupt onset, which is called a ``fast
trigger.'' However, a satisfactory explanation of this fast trigger is
not yet known. Another unresolved problem is the possibility of a
partial sawtooth collapse, where the $m=1$ mode saturates before the
reconnection of the core is completed, so only an annular (off-axis)
region undergoes mixing \cite{Hastie98}.

During the evolution of a tokamak plasma subject to sawtooth
relaxation oscillations, multiple $q=1$ resonant surfaces may arise
temporarily \cite{Aydemir89}. When this occurs in configurations with
a hollow current profile ($q_0 > 1$ with $q_0$ being the $q$ at the
magnetic axis) $q=1$ double tearing modes (DTMs) can become unstable
\cite{White77, Parail80, Pfeiffer85} whereas, for a centrally peaked
current profile ($q_0 < 1$), $q=1$ triple tearing modes (TTMs) may
arise \cite{Pfeiffer85}.

In this letter, it is demonstrated that some of the phenomena
associated with the sawtooth oscillations in tokamaks can be explained
by the nonlinear evolution of DTMs or TTMs. Using nonlinear numerical
simulations, it is shown that, in the presence of multiple $q=1$
resonant surfaces, rapidly growing high-$m$ DTMs or TTMs can enhance
the growth of the $m=1$ mode and later generate electromagnetic
turbulence in the annular region surrounded by the $q=1$ resonant
surfaces. Based on these observations, it is shown that the fast
trigger of a sawtooth crash as well as precursor-free partial
collapses during sawtooth relaxation oscillations can be accounted for
by the nonlinear evolution of TTMs.

The set of equations we use is the reduced magnetohydrodynamic (RMHD)
equation in the zero-beta limit in a cylindrical geometry.
%\cite{Strauss76}
In normalized form the RMHD model can be written as
\begin{eqnarray}
\partial_t\psi &=& \left[\psi,\phi\right] - \partial_\zeta\phi -
\Shp^{-1}\left(\hat{\eta}j - E_0\right)
\label{eq:rmhd1}
\\
\partial_t u &=& \left[u,\phi\right] + \left[j,\psi\right] +
\partial_\zeta j + \nu\nabla_\perp^2 u,
\label{eq:rmhd2}
\end{eqnarray}

\noindent where $\psi$ is the magnetic flux function, $\phi$ the
stream function (electrostatic potential), $j =
-\nabla_\perp^2\psi$ the axial current density and $u =
\nabla_\perp^2\phi$ the vorticity, essentially following the
standard notation (cf., e.g., Ref.~\cite{NishikawaWakatani}). The
time $t$ is normalized by the poloidal Alfv\'{e}n time $\tHp$
(time scale for dynamics in an ideal magnetized plasma) and the
radial coordinate by the minor radius $a$ of the plasma column.
The resistivity profile is given by $\hat{\eta}(r) =
j(r=0,t=0)/j(r,t=0)$. As to the Lundquist number $\Shp = \tau_{\rm
R}/\tHp$ [where $\tau_{\rm R} = \mu_0/(\eta_0 a^2)$ is the
resistive time scale and $\eta_0$ the resistivity at $r=0$] and
viscosity $\nu$ (normalized by $a^2/\tHp$), $\Shp = \nu^{-1} =
10^6$ is used, unless stated otherwise. The constant source term
$E_0$ compensates the resistive dissipation of the equilibrium
current. Using a quasi-spectral code with a finite-difference
radial mesh, single helicity ($h=m/n=1$, $0 \leq m \leq 127$)
nonlinear simulations of Eq.~(\ref{eq:rmhd1}) and (\ref{eq:rmhd2})
were performed with an initial condition corresponding to a
flow-less equilibrium with perturbed flux function
($\psi_{m>0}(t=0) = 10^{-11}$, $m=n$). An ideally conducting wall
was taken as the boundary condition. Although the RMHD model
encompasses only part of the physical effects involved in sawtooth
oscillations (cf., e.g.,
Refs.~\cite{WangBhatt95,Porcelli96,Hastie98}), it suffices for the
present purpose of following the fundamental dynamics of DTMs and
TTMs.

\begin{figure}
[tbp]
\centering
\includegraphics[height=6.0cm,width=8.0cm]% aspect ratio: 1.333
{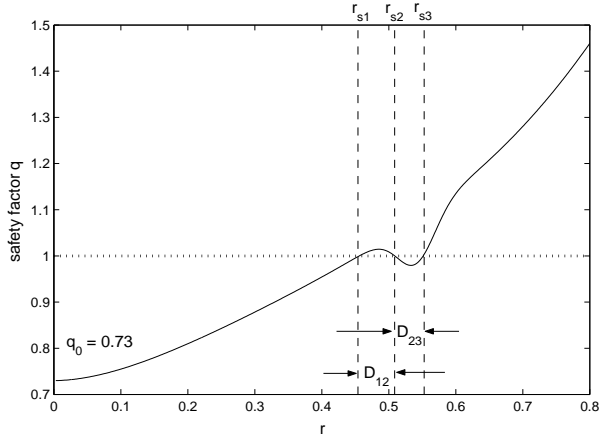}%
\caption{A safety factor profile $q(r)$ unstable to TTMs. Vertical
lines (dashed) indicate the locations of the $q=1$ resonant surfaces,
$\rsa < \rsb < \rsc$, and $q=1$ is indicated by a horizontal (dotted)
line.}
\label{fig:equlib_3tm}%
\end{figure}

\begin{figure}
[tbp]
\centering
\includegraphics[height=6.0cm,width=8.0cm]% aspect ratio: 1.333
{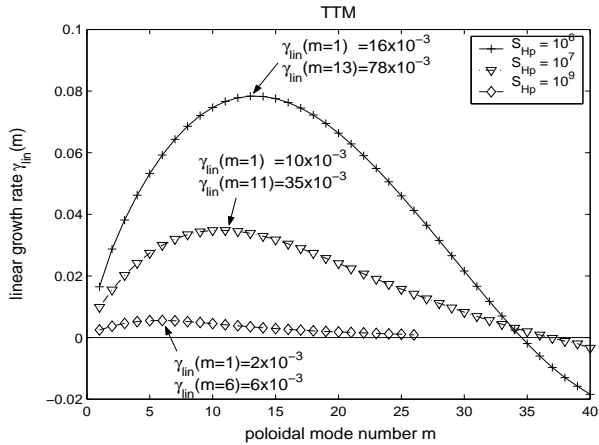}%
\caption{Linear growth rate spectra of TTMs for the $q$ profile
given in Fig.~\protect\ref{fig:equlib_3tm} for the Lundquist
numbers $\Shp = 10^6, 10^7$ and $10^9$ and constant Prandtl number
$Pr = \Shp\nu = 1$.}
\label{fig:spec-eta_3tm}%
\end{figure}

\begin{figure}
[tbp]
\centering
\includegraphics[height=6.0cm,width=8.0cm]% aspect ratio: 1.333
{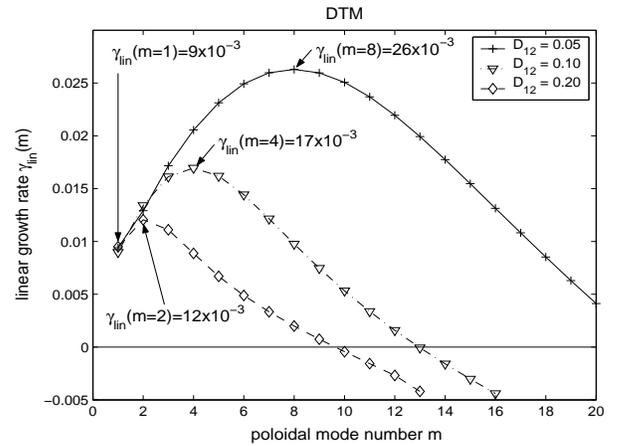}%
\caption{Linear growth rate spectra of DTMs for $q$ profiles with two
$q=1$ resonant surfaces located a distance $D_{12}$ apart, obtained
with $\Shp = \nu^{-1} = 10^6$. In the three cases shown, $D_{12} =
0.05$, $0.1$ and $0.2$, respectively.}
\label{fig:spec-dr_2tm}%
\end{figure}

\begin{figure}
[tbp] \centering
\includegraphics[height=6.0cm,width=8.0cm]% aspect ratio: 1.333
{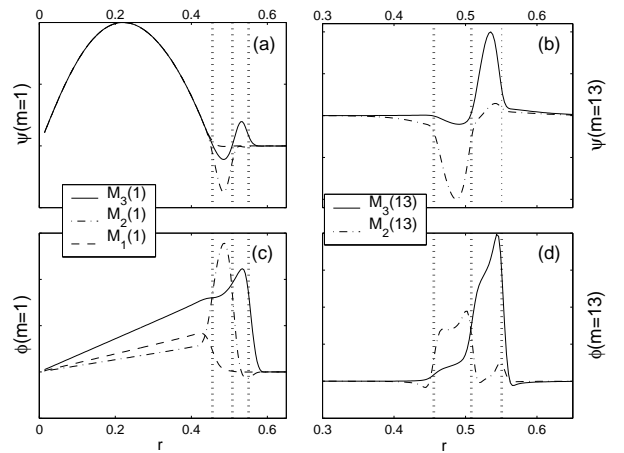}%
\caption{Radial structure of linearly unstable TTM eigenmodes for
(a) the $m=1$ modes of $\psi$, (b) the $m=13$ modes of $\psi$, (c)
the $m=1$ modes of $\phi$, and (d) the $m=13$ modes of $\phi$
obtained for $\Shp = \nu^{-1} = 10^6$. For a given $m$, the
eigenmode extending to the resonant radius $r_{{\rm s}i}$ is
denoted by $M_i(m)$. Vertical lines (dotted) indicate resonant
radii.}
\label{fig:mstruc_m1-13}%
\end{figure}

The important new feature of \emph{linear} instability that is
addressed here is the fact that configurations with multiple $q=1$
resonant surfaces in general possess a \emph{broad spectrum} of
linearly unstable modes. Moreover, the fastest growing mode often
has a poloidal mode number $m \sim {\mathcal O}(10)$. To show
this, the $q$ profile shown in Fig.~\ref{fig:equlib_3tm} is
employed, where three $q=1$ resonant surfaces are present, at the
radii $\rsa < \rsb < \rsc$. By evolving the linearized RMHD
equations in time, spectra of linear growth rates (i.e.,
dispersion relations) $\gamma_{\rm lin}(m)$ were obtained as
functions of $m$, as plotted in Fig.~\ref{fig:spec-eta_3tm} for
Lundquist numbers $\Shp = 10^6$, $10^7$ and $10^9$. Hereby the
Prandtl number $Pr = \Shp\nu$ has been kept equal to unity.
Clearly, a variation of $\Shp$ (while $Pr = 1$) retains the
broadness of the spectrum and $\gamma_{\rm max} = {\rm
Max}\{\gamma_{\rm lin}(m)\}$ is located at $m > 1$ in all cases.

\begin{figure}
[tbp] \centering
\includegraphics[height=6.0cm,width=8.0cm]% aspect ratio: 1.333
{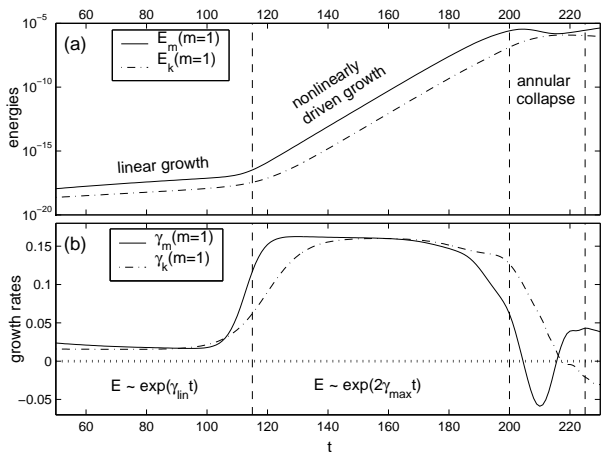} \caption{Evolution of the $m=1$ TTM for the
$q$ profile given in Fig.~\protect\ref{fig:equlib_3tm}. The
perturbed kinetic and magnetic energies, denoted by $E_{\rm k}$
and $E_{\rm m}$ are plotted in (a). The growth rates $\gamma_{\rm
k}$ and $\gamma_{\rm m}$ given in (b) are estimated from $E_{\rm
k}$ and $E_{\rm m}$, and $\gamma_{\rm max} \equiv {\rm
Max}\{\gamma_{\rm lin}(m)\}$ is the maximum growth rate in the
spectrum in Fig.~\protect\ref{fig:spec-eta_3tm} for $\Shp = 10^6$.
The system reaches the fully nonlinear saturation in the phase
denoted as ``annular collapse,'' i.e., $t > 200$.}
\label{fig:E-g_3tm}%
\end{figure}

\begin{figure}
[tbp] \centering
\includegraphics[height=7.5cm,width=7.5cm]% aspect ratio: 1.0
{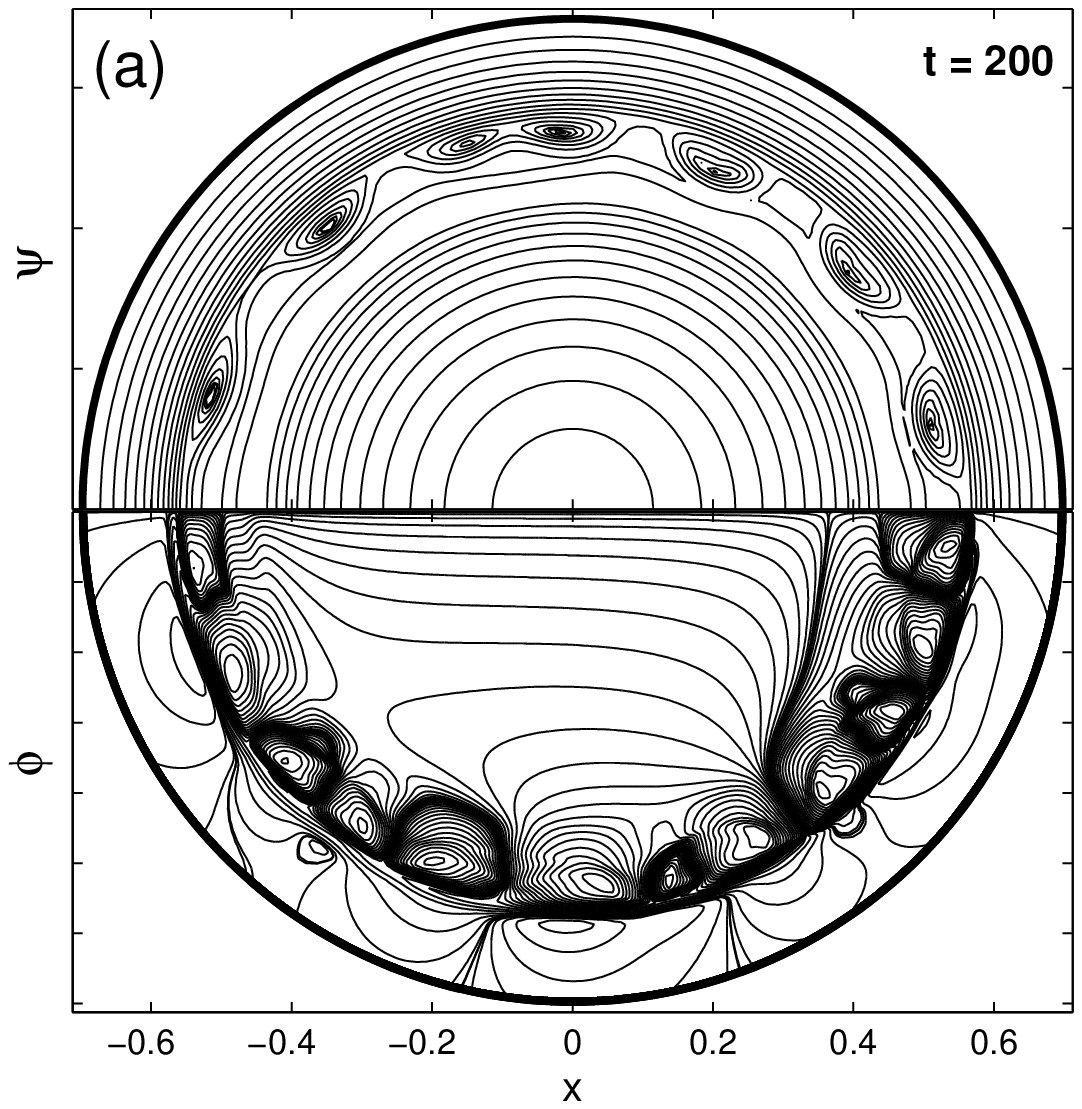}
\includegraphics[height=7.5cm,width=7.5cm]% aspect ratio: 1.0
{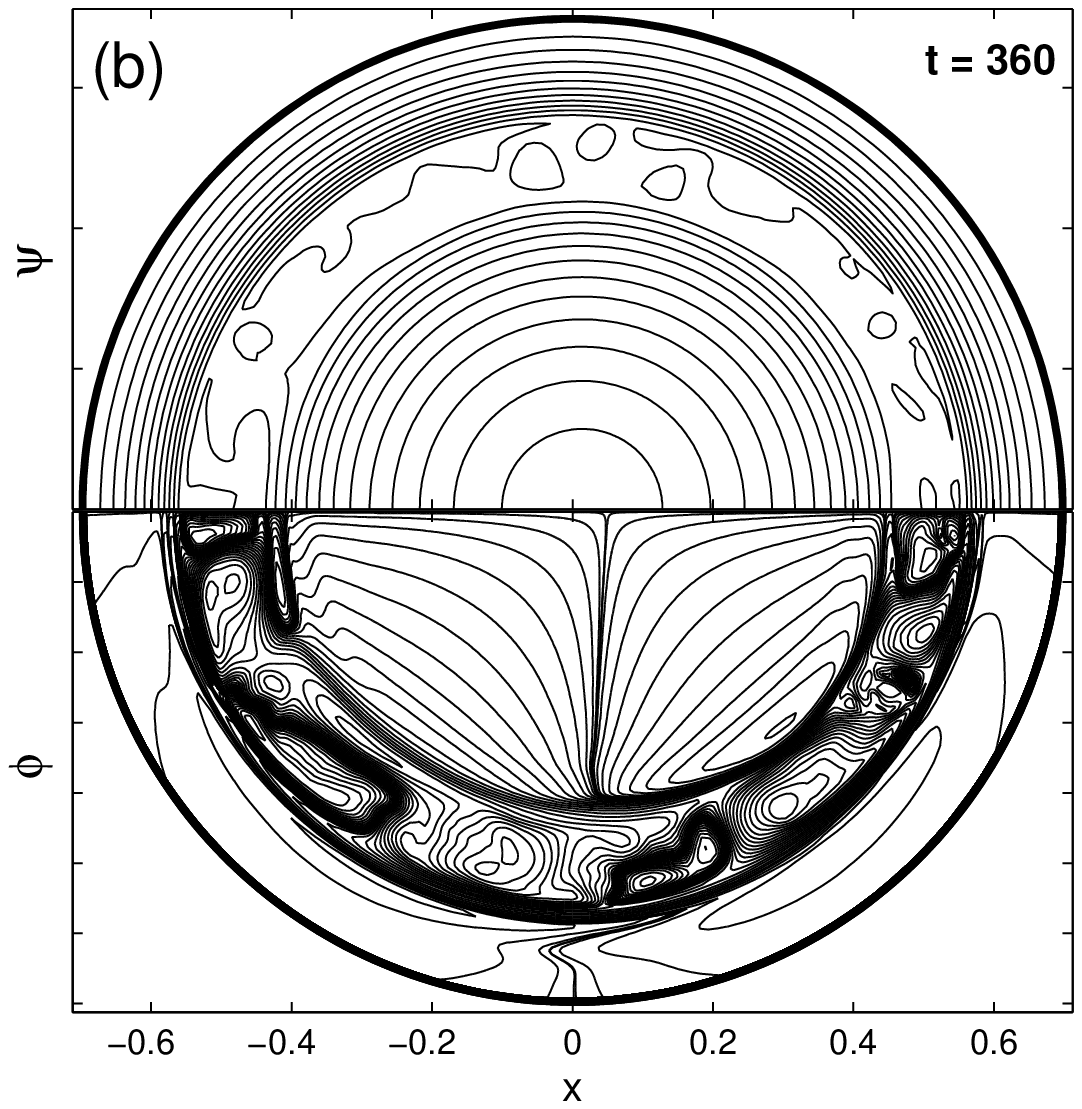} \caption{Upper and lower halves of
$\psi$ (top) and $\phi$ (bottom) contour lines in the poloidal
cross-section during the annular collapse (a) and after the $q$
profile was annularly flattened (b).}
\label{fig:snaps_turb}%
\end{figure}

The dependence of the spectrum $\gamma_{\rm lin}(m)$ on the distance
between the $q=1$ surfaces is most easily investigated by considering
DTM configurations where two $q=1$ resonant surfaces are present,
located at radii $\rsa$ and $\rsb$, a distance $D_{12} = \rsb - \rsa$
apart. In Fig.~\ref{fig:spec-dr_2tm} the DTM growth rate spectra for
profiles with $D_{12} = 0.05$, $0.1$ and $0.2$ are shown. While
varying $D_{12}$, the local magnetic shears at the resonant radii,
$s_1 = s(\rsa)$ and $s_2 = s(\rsb)$, were not changed. It can be seen
that the narrower the inter-resonance region becomes, the more the
poloidal mode number of the fastest growing mode shifts to larger
values of $m$, and $\gamma_{\rm max}$ increases. On the other hand,
the growth rate of the $m=1$ mode hardly depends on $D_{12}$ and it
becomes the fastest growing mode for sufficiently large values of
$D_{12}$. Similar results are found for TTMs, whereby it is noted that
TTMs tend to peak at higher $m$ with higher growth rates than DTMs
with similar $D_{ij}$. The spectra shown in
Figs.~\ref{fig:spec-eta_3tm} and \ref{fig:spec-dr_2tm} contrast with
that for the single tearing modes (STM), for which the $m=1$ mode is
dominant and the modes with higher $m$ are usually linearly
stable. Note that the broad spectra of tokamak TTMs that were obtained
here are similar to those of tearing modes obtained by Dahlburg and
Karpen \cite{Dahlburg95} for triple current sheets (TCS) in slab
geometry as a model for adjoining helmet streamers in the solar
corona.

It must be emphasized that the dispersion curves in
Fig.~\ref{fig:spec-eta_3tm} show only the growth rate of the most
unstable mode for each $m$. However in general, for a given $m$
there are up to three unstable TTM eigenmodes, each associated
with a resonant surface. To illustrate this, the radial structure
of the eigenmodes for $m=1$ and $m=13$ are shown in
Fig.~\ref{fig:mstruc_m1-13}, which were obtained by solving the
eigenvalue problem for the linearized Eqs.~(\ref{eq:rmhd1}) and
(\ref{eq:rmhd2}). Here $M_1(m)$ denotes the eigenmode with the
poloidal mode number $m$ that extends only to the innermost
resonant surface $\rsa$. Note that $M_1(1)$ has the same mode
structure as an STM. Similarly, $M_2(m)$ and $M_3(m)$ denote the
eigenmodes that extend to $\rsb$ and $\rsc$, respectively. It is
also noted that, for $m=13$, $M_1$ (not shown) is stable, as can
be expected from the linear stability of STMs with higher $m$.
Similar eigenmode structures are found for DTMs, and indeed
similar instability characteristics are also expected for $q$
profiles with more than three $q=1$ resonant surfaces.

After perturbing a large number of unstable modes at random
poloidal angles, the $m=1$ mode evolves as shown in
Fig.~\ref{fig:E-g_3tm}. The most remarkable feature here is the
presence of a phase of \emph{nonlinearly driven growth}. There,
the energy of the $m=1$ perturbed mode grows exponentially as
$\exp(\gamma_{\rm drive}t)$. In this example, $\gamma_{\rm drive}
\approx 0.16$, i.e., the $m=1$ mode grows at a rate which is one
order of magnitude larger than its linear growth rate $\gamma_{\rm
lin}(m=1) = 16\times 10^{-3}$. The nonlinear growth rate
$\gamma_{\rm drive}$ approximately equals twice the maximum growth
rate in the spectrum (Fig.~\ref{fig:spec-eta_3tm}), $\gamma_{\rm
max} = 0.08$, because it results from the nonlinear coupling of
$m$ and $m+1$ mode pairs.

The growth rates shown in Fig.~\ref{fig:spec-eta_3tm} belong
exclusively to $M_3(m)$ modes, i.e., the modes extending to the
outermost resonant surface at $r=\rsc$, since it was assumed here that
$s_1 < s_3$ ($s_1 = 0.35$, $s_2 = -0.56$, $s_3 =1.20$). On the
other hand, for a $q$ profile with $s_1 > s_3$, the eigenmode
$M_1(m=1)$ has a higher growth rate than $M_3(m=1)$. However, also in
this case, the highest growth rate among all $m$ modes, $\gamma_{\rm
max}$, is typically several times larger than $\gamma_{\rm lin}(m=1)$
and therefore the nonlinearly driven growth of the $m=1$ mode still
exceeds its linear growth. It is concluded that the rapid nonlinear
growth of TTMs occurs in a wide range of TTM $q$ profiles. Since the
sawtooth crash is generally considered to be triggered by the onset of
an $m=1$ mode \cite{Porcelli96}, the results presented here suggest
that the nonlinear growth of TTMs (or DTMs for a hollow current
profile) is one of the possible mechanisms for experimentally observed
abrupt sawtooth crashes, i.e., the fast trigger. Let us note that
coupled tearing modes on two nearby resonant surfaces can exhibit
exponential non-Rutherford growth to saturation even for $q>1$
\cite{White77,Ishii02}.

Finally, the fully nonlinear regime of Fig.~\ref{fig:E-g_3tm} is
discussed. At about $t=200$ electromagnetic turbulence starts to
develop in the whole inter-resonance region $\rsa<r<\rsc$, as can
be seen in Fig.~\ref{fig:snaps_turb} (a). The fluctuations flatten
the $q$ profile through simultaneous magnetic reconnection in the
\emph{whole} inter-resonance region. This \emph{partial collapse}
(i.e., not involving the core) is essentially completed around
$t=220\sim 230$. A contour plot showing the situation after the
$q$ profile was annularly flattened is given in
Fig.~\ref{fig:snaps_turb} (b). An important property of an annular
collapse as the one shown in Fig.~\ref{fig:snaps_turb} is that no
$m=1$ islands needs to form and therefore the displacement of the
core plasma can be rather small.

This phenomenon is similar to the experimentally observed off-axis
temperature collapses without precursor oscillations. For example,
Edwards \textit{et al.} \cite{Edwards86} reported JET experiments
where rapid sawtooth crashes without evident precursor oscillations,
but preceded by a partial crash in an off-axis region, were
observed.

A possible explanation for such phenomena was proposed by Buratti
\textit{et al.} in terms of a ``purely growing precursor''
\cite{Buratti03}. However, the results in
Fig.~\ref{fig:snaps_turb} show that, if a $q$ profile with
relatively small inter-resonance distances $D_{12}$ and $D_{23}$
is formed even temporarily, rapidly growing nonlinear TTMs can
also cause such partial crashes without precursors.

Precursor-free partial collapses during the ramp phase of compound
sawtooth oscillations were also observed in tokamak discharges
with hollow $q$ profiles
\cite{Taylor86,Campbell86,Kim86,Ishida88}, where $q=1$ DTMs can be
expected. In analogy with the TTM case discussed above, our
results indicate that nonlinear growth of $q=1$ DTMs can lead to
such a partial collapse. In cases where the growth rate spectrum
peaks at lower $m$ (e.g., due to larger $D_{ij}$ or higher
$\Shp$), we have also observed partial collapses (i.e., collapses
in the inter-resonance region) after which the core displacement
continues to grow, which is similar to experimental observations
made in JET, given in Fig.~4 (A)-(C) in Ref.~\cite{Edwards86}.

In summary, our numerical simulations have shown that the simultaneous
excitation of unstable $q=1$ TTMs with high $m$ and their subsequent
nonlinear interactions lead to a rapid onset of the $m=1$ triple
tearing mode, which qualitatively depicts the fast triggering of
sawtooth crashes observed in tokamak experiments. Similar behavior has
also been found in simulations of $q=1$ DTMs. If more than three
$q=1$ resonant surfaces are formed in a tokamak discharge, we also
expect similar multiple tearing modes that grow rapidly with high
poloidal mode numbers. We have also presented a scenario, where the
nonlinear evolution of many unstable TTMs leads to a partial collapse
of a sawtooth without being preceded by an $m=1$ precursor. Similar
phenomena were observed during compound sawtooth oscillations in
several experiments
\cite{Edwards86,Taylor86,Campbell86,Kim86,Ishida88}.

A.B. would like to thank Y. Kishimoto, Y. Nakamura and M. Yagi for
valuable discussions. S.B. acknowledges the Graduate School of
Energy Science for its support and hospitality. This work is
partially supported by the 21st Century COE Program at Kyoto
University.

%===========================================

\end{document}